\documentclass[pra,aps,twocolumn]{revtex4}

\usepackage{amssymb}
\usepackage{amsmath}
\usepackage{epsfig}

\begin{document}
\title{Asymmetric quantum telecloning of multiqubit states}
\author{Lin Chen}
\author{Yi-Xin Chen}
\affiliation{Zhejiang Insitute of Modern Physics, Zhejiang
University, Hangzhou 310027, China}

\begin{abstract}
We propose a scheme of 1$\rightarrow$2 optimal universal
asymmetric quantum telecloning of pure multiqubit states. In
particular, we first investigate the asymmetric telecloning of
arbitrary 2-qubit states and then extend it to the case of
multiqubit system. Many figures of merit for the telecloning
process are checked, including the entanglement of the quantum
channel and fidelities of the clones. Our scheme can be used for
the 1$\rightarrow$4 universal telecloning of mixed multiqubit
states.
\end{abstract}
\maketitle

\section{Introduction}
One of the most essential differences between classical and
quantum-information theory (QIT) is the no cloning theorem
\cite{Zurek,Scarani}. It forbids the perfect cloning of
arbitrarily given quantum state, in both pure and mixed cases. It
is then natural to ask how well one can copy quantum states, i.e.,
with the highest fidelity. This problem was firstly addressed by
Buzek and Hillery \cite{Buzek1}, whose scheme proved to be optimal
by \cite{Bruss}. The Buzek-Hillery theory actually exhibits a
universal symmetric 1$\rightarrow$2 quantum cloning machine (QCM),
which exports two identical clones closest to the input pure qubit
state with a constant fidelity. The related work in past years has
established the $N \rightarrow M$ universal symmetric QCM for both
qubits \cite{Gisin} and qudits \cite{Werner,Murao1} (transforming
$N$ identical input states into $M>N$ identical output copies), as
well as the continuous-variable systems \cite{Cerf1}.
Correspondingly, the $N \rightarrow M$ asymmetric QCM generates
$M$ output states with different fidelities from $N$ input copies
\cite{Niu,Cerf2,Braunstein,Iblisdir}. Some experimental progress
on quantum cloning has also been made \cite{Ricci}.

The essentiality of quantum cloning is to broadcast information to
certain distributed objects, so it is regarded as a widely useful
quantum-information transmission, e.g., the eavesdropping on
implementation of quantum key distribution \cite{Curty}. It is
well-known that quantum teleportation \cite{Bennett,Bouwmeester}
is the most effective technique for remotely broadcasting
information. Murao \textit{et al.} \cite{Murao2} has advanced the
$1 \rightarrow M$ quantum telecloning which combines the tricks of
both quantum teleportation and cloning. In this scheme, the sender
Alice holds an unknown input state and she previously shares an
entangled state with $M$ receivers, which resembles the scenario
of quantum teleportation. The object is to duplicate the input at
the location of every receiver as well as possible, since the
no-cloning theorem precludes the faithful copy of unknown quantum
state. Similarly, there exist symmetric and asymmetric quantum
telecloning with identical and different fidelities of the clones
respectively. The technique of symmetric telecloning has been
extended to the case of $N \rightarrow M$ for qubit states
\cite{Dur} and $1 \rightarrow M$ for qudits \cite{Murao1}, while
the $1 \rightarrow 2$ universal optimal asymmetric telecloning was
realized by \cite{Murao1,Ghiu1}. Of all these traditional schemes,
the input states are restricted in the local scenario, namely the
sender can arbitrarily perform the unitary operation on its
system. This is no longer correct when the input states are
entangled, and some primary investigation for entanglement cloning
has been made recently \cite{Cerf3}. However, they merely found
out the condition on which a universal QCM can be optimal for the
input of maximally entangled states, and this problem proves
exceedingly difficult. As the entanglement plays the essential
role in QIT, it is significant to explore the cloning and
telecloning of entanglement. Unlike the broadcasting of entangled
states \cite{Plenio}, entanglement telecloning is optimal in the
sense that it achieves the best fidelity as those of universal
symmetric QCMs for qudits and so on. Recently, \cite{Ghiu2}
proposed the scheme of telecloning for the entangled inputs
$\left|\psi\right\rangle=\alpha\left|00\right\rangle+\beta\left|11\right\rangle$,
which is a small set of the two-qubit states (we refer to
$\left|j\right\rangle,j=0,1,...$ as the computational basis in
this paper, see below). It is then interesting to extend this
scheme to the case of general two-qubit inputs. However, we doubt
that whether such a scheme can be universal for any input, namely
with a constant fidelity. If so, can it reach the optimal fidelity
of the universal QCMs such as Werner's bound \cite{Werner} ?
Furthermore, as the extensive use of multipartite entanglement, it
is important to explore the telecloning of multiqubit states.

In the present work, we propose a scheme of 1$\rightarrow$2
optimal universal asymmetric quantum telecloning of pure
multiqubit states, by virtue of the Heisenberg QCM in
\cite{Cerf2,Ghiu1}. Based on this motivation, we firstly
investigate the asymmetric telecloning of arbitrary two-qubit
states and compare the achievable fidelity with the existing
universal QCMs. The required entanglement in this scheme is shown
to be optimal for the 4-dimensional input states. We also
explicitly prove that the telecloner never creates more
entanglement than that contained in the input qubits. Furthermore,
we extend the above scheme to the case of multiqubit inputs. Thus
we have realized for the first time the universal telecloning of
arbitrarily nonlocal multiqubit states. As a $d-$level system can
be composed of many qubits, one can hence optimally teleclone any
state by our scheme. An important application of this technique is
to perform the 1$\rightarrow$4 telecloning of mixed multiqubit
states, which is a greatly puzzling problem in QIT \cite{Scarani}.
Strikingly, we find that such a scheme can be realized with a
higher fidelity than that of the optimal telecloning of pure
states. We thus make resultful progress to get insight into the
field of telecloning of mixed states.

The paper is organized as follows. In Sec. II we present the
explicit protocol for the case of 2-qubit input states, and
investigate the properties of the telecloning process. In Sec. III
we extend it to the case of multiqubit inputs and apply it to the
1$\rightarrow$4 telecloning of arbitrary mixed states. We present
our conclusion in Sec. IV.

\section{Optimal universal 1$\rightarrow$2 telecloning of 2-qubit states}

As shown in \cite{Bennett,Murao1}, either of quantum teleportation
and telecloning requires an unknown input state, which is to be
reconstructed in several remotely distributed places. In the
present situation, the input state shared by two parties $A_1,A_2$
has the following form,
\begin{equation}
\left|\psi\right\rangle_{A_1A_2}=\alpha_0\left|00\right\rangle+\alpha_1\left|01\right\rangle+
\alpha_2\left|10\right\rangle+\alpha_3\left|11\right\rangle,
\end{equation}
where $\alpha_i\in\mathcal{C},\forall i$, and
$|\alpha_0|^2+|\alpha_1|^2+|\alpha_2|^2+|\alpha_3|^2=1$ by the
normalization condition. The aim of the telecloning is to
respectively transimit two copies of this state to two groups of
receivers $B_1,B_2$ and $C_1,C_2$ with the highest fidelity, where
every party can only operate on their states locally with the help
of classical communication. To find out the appropriate quantum
channel between senders and receivers, we recall the optimal
universal asymmetric Heisenberg QCM \cite{Cerf2,Ghiu1},
\begin{eqnarray}
U\left|j\right\rangle_B\left|00\right\rangle_{C,Anc}&=&\frac{1}{\sqrt{1+(d-1)(p^2+q^2)}}\nonumber\\
&\times&\Big(\left|j\right\rangle\left|j\right\rangle\left|j\right\rangle+
p\sum^{d-1}_{r=1}\left|j\right\rangle\left|\overline{j+r}\right\rangle\left|\overline{j+r}\right\rangle\nonumber\\
&+&
q\sum^{d-1}_{r=1}\left|\overline{j+r}\right\rangle\left|j\right\rangle\left|\overline{j+r}\right\rangle\Big),\nonumber\\
0\leq j \leq d-1,
\end{eqnarray}
which indeed represents series of QCMs by altering $j$. Here,
$\overline{j+r}=j+r$ modulo $d$ and $d$ is the dimension of the
input state. The real constants $p,q$ satisfy $p+q=1$ and their
concrete meaning is to generate a universal QCM and to keep the
optimality of it, so they can be properly defined previously. By
superposition of the QCMs in expression (2), one can set an
arbitrary state of system $B$ as input to obtain two clones at
systems $B,C$ respectively, and the third qudit is the ancilla.

In a $d-$dimension Hilbert space, the computational basis can be
expressed as a composition of the qubits, e.g., when $d=4$ we
denote that
$\left|0\right\rangle\rightarrow\left|00\right\rangle$,
$\left|1\right\rangle\rightarrow\left|01\right\rangle$,
$\left|2\right\rangle\rightarrow\left|10\right\rangle$ and
$\left|3\right\rangle\rightarrow\left|11\right\rangle$. For
simplicity, let $\left|\eta_j\right\rangle\equiv
U\left|j\right\rangle_B\left|00\right\rangle_{C,Anc},j=0,1,...$.
In our scheme, either of $B,C$ and the ancilla should be a
composite system of two separated qubits, whose dimension is at
most $d=4$. Concretely, we write out $\left|\eta_j\right\rangle$'s
from expression (2),
\begin{eqnarray}
\left|\eta_0\right\rangle
&=&[1+3(p^2+q^2)]^{-1/2}\big(\left|00\right\rangle\left|00\right\rangle\left|00\right\rangle\nonumber\\
&+&p\left|00\right\rangle\left|01\right\rangle\left|01\right\rangle
+p\left|00\right\rangle\left|10\right\rangle\left|10\right\rangle
+p\left|00\right\rangle\left|11\right\rangle\left|11\right\rangle\nonumber\\
&+&q\left|01\right\rangle\left|00\right\rangle\left|01\right\rangle
+q\left|10\right\rangle\left|00\right\rangle\left|10\right\rangle
+q\left|11\right\rangle\left|00\right\rangle\left|11\right\rangle\big)\nonumber\\
\left|\eta_1\right\rangle
&=&[1+3(p^2+q^2)]^{-1/2}\big(\left|01\right\rangle\left|01\right\rangle\left|01\right\rangle\nonumber\\
&+&p\left|01\right\rangle\left|00\right\rangle\left|00\right\rangle
+p\left|01\right\rangle\left|10\right\rangle\left|10\right\rangle
+p\left|01\right\rangle\left|11\right\rangle\left|11\right\rangle\nonumber\\
&+&q\left|00\right\rangle\left|01\right\rangle\left|00\right\rangle
+q\left|10\right\rangle\left|01\right\rangle\left|10\right\rangle
+q\left|11\right\rangle\left|01\right\rangle\left|11\right\rangle\big)\nonumber\\
\left|\eta_2\right\rangle
&=&[1+3(p^2+q^2)]^{-1/2}\big(\left|10\right\rangle\left|10\right\rangle\left|10\right\rangle\nonumber\\
&+&p\left|10\right\rangle\left|00\right\rangle\left|00\right\rangle
+p\left|10\right\rangle\left|01\right\rangle\left|01\right\rangle
+p\left|10\right\rangle\left|11\right\rangle\left|11\right\rangle\nonumber\\
&+&q\left|00\right\rangle\left|10\right\rangle\left|00\right\rangle
+q\left|01\right\rangle\left|10\right\rangle\left|01\right\rangle
+q\left|11\right\rangle\left|10\right\rangle\left|11\right\rangle\big)\nonumber\\
\left|\eta_3\right\rangle
&=&[1+3(p^2+q^2)]^{-1/2}\big(\left|11\right\rangle\left|11\right\rangle\left|11\right\rangle\nonumber\\
&+&p\left|11\right\rangle\left|00\right\rangle\left|00\right\rangle
+p\left|11\right\rangle\left|01\right\rangle\left|01\right\rangle
+p\left|11\right\rangle\left|10\right\rangle\left|10\right\rangle\nonumber\\
&+&q\left|00\right\rangle\left|11\right\rangle\left|00\right\rangle
+q\left|01\right\rangle\left|11\right\rangle\left|01\right\rangle
+q\left|10\right\rangle\left|11\right\rangle\left|10\right\rangle\big)\nonumber\\.
\end{eqnarray}
Then we propose that the quantum channel shared by all parties is
\begin{eqnarray}
\left|\Omega\right\rangle_{A^{\prime}_1A^{\prime}_2B_1B_2C_1C_2a_1a_2}&=&
\frac12\big(\left|00\right\rangle_{A^{\prime}_1A^{\prime}_2}\left|\eta_0\right\rangle_{B_1B_2C_1C_2a_1a_2}\nonumber\\
&+&\left|01\right\rangle_{A^{\prime}_1A^{\prime}_2}\left|\eta_1\right\rangle_{B_1B_2C_1C_2a_1a_2}\nonumber\\
&+&\left|10\right\rangle_{A^{\prime}_1A^{\prime}_2}\left|\eta_2\right\rangle_{B_1B_2C_1C_2a_1a_2}\nonumber\\
&+&\left|11\right\rangle_{A^{\prime}_1A^{\prime}_2}\left|\eta_3\right\rangle_{B_1B_2C_1C_2a_1a_2}\big),
\end{eqnarray}
where $A^{\prime}_1$ and $A^{\prime}_2$ are two particles
belonging to the senders $A_1$ and $A_2$ respectively. Notice the
two ancillas $a_1,a_2$ are held by some separated observers. The
ancilla particles are necessary for the Heisenberg QCM, otherwise
it cannot reach the optimal fidelity \cite{Cerf2}. Although the
ancillas do not play the role of clones, we will see that they
actually join the realization of optimal telecloning of
entanglement. For example, there are some useful relations with
respect to the states $\left|\eta_i\right\rangle$'s, which
involves all the participants in the system
\begin{eqnarray}
\sigma_{zB_1}\otimes\sigma_{zC_1}\otimes\sigma_{za_1}\left|\eta_i\right\rangle=\left|\eta_i\right\rangle,i=0,1\\
\sigma_{zB_1}\otimes\sigma_{zC_1}\otimes\sigma_{za_1}\left|\eta_i\right\rangle=-\left|\eta_i\right\rangle,i=2,3\\
\sigma_{zB_2}\otimes\sigma_{zC_2}\otimes\sigma_{za_2}\left|\eta_i\right\rangle=\left|\eta_i\right\rangle,i=0,2\\
\sigma_{zB_2}\otimes\sigma_{zC_2}\otimes\sigma_{za_2}\left|\eta_i\right\rangle=-\left|\eta_i\right\rangle,i=1,3\\
\sigma_{xB_1}\otimes\sigma_{xC_1}\otimes\sigma_{xa_1}\left|\eta_0\right\rangle=\left|\eta_2\right\rangle,\\
\sigma_{xB_1}\otimes\sigma_{xC_1}\otimes\sigma_{xa_1}\left|\eta_1\right\rangle=\left|\eta_3\right\rangle,\\
\sigma_{xB_2}\otimes\sigma_{xC_2}\otimes\sigma_{xa_2}\left|\eta_0\right\rangle=\left|\eta_1\right\rangle,\\
\sigma_{xB_2}\otimes\sigma_{xC_2}\otimes\sigma_{xa_2}\left|\eta_2\right\rangle=\left|\eta_3\right\rangle.
\end{eqnarray}
These equations can be easily checked by using of the expressions
of $\left|\eta_i\right\rangle$'s. Specially, the first four
equations represent the change of the sign while the last four
represent the change between the states
$\left|\eta_i\right\rangle$'s. We thus call them
parity-transformation and state-transformation respectively.

In what follows we show how to carry out the universal optimal
$1\rightarrow2$ telecloning of the two-qubit state
$\left|\psi\right\rangle_{A_1A_2}=\alpha_0\left|00\right\rangle+\alpha_1\left|01\right\rangle+
\alpha_2\left|10\right\rangle+\alpha_3\left|11\right\rangle$. The
whole system is in the state
\begin{eqnarray}
\left|\Psi\right\rangle_{tot}=\left|\psi\right\rangle_{A_1A_2}\otimes
\left|\Omega\right\rangle_{A^{\prime}_1A^{\prime}_2B_1B_2C_1C_2a_1a_2}=\nonumber\\
\frac{\alpha_0}{2}
(\left|00\right\rangle_{A_1A^{\prime}_1}\left|00\right\rangle_{A_2A^{\prime}_2}\left|\eta_0\right\rangle+
\left|00\right\rangle_{A_1A^{\prime}_1}\left|01\right\rangle_{A_2A^{\prime}_2}\left|\eta_1\right\rangle+\nonumber\\
\left|01\right\rangle_{A_1A^{\prime}_1}\left|00\right\rangle_{A_2A^{\prime}_2}\left|\eta_2\right\rangle+
\left|01\right\rangle_{A_1A^{\prime}_1}\left|01\right\rangle_{A_2A^{\prime}_2}\left|\eta_3\right\rangle)\nonumber\\
+\frac{\alpha_1}{2}
(\left|00\right\rangle_{A_1A^{\prime}_1}\left|10\right\rangle_{A_2A^{\prime}_2}\left|\eta_0\right\rangle+
\left|00\right\rangle_{A_1A^{\prime}_1}\left|11\right\rangle_{A_2A^{\prime}_2}\left|\eta_1\right\rangle+\nonumber\\
\left|01\right\rangle_{A_1A^{\prime}_1}\left|10\right\rangle_{A_2A^{\prime}_2}\left|\eta_2\right\rangle+
\left|01\right\rangle_{A_1A^{\prime}_1}\left|11\right\rangle_{A_2A^{\prime}_2}\left|\eta_3\right\rangle)\nonumber\\
+\frac{\alpha_2}{2}
(\left|10\right\rangle_{A_1A^{\prime}_1}\left|00\right\rangle_{A_2A^{\prime}_2}\left|\eta_0\right\rangle+
\left|10\right\rangle_{A_1A^{\prime}_1}\left|01\right\rangle_{A_2A^{\prime}_2}\left|\eta_1\right\rangle+\nonumber\\
\left|11\right\rangle_{A_1A^{\prime}_1}\left|00\right\rangle_{A_2A^{\prime}_2}\left|\eta_2\right\rangle+
\left|11\right\rangle_{A_1A^{\prime}_1}\left|01\right\rangle_{A_2A^{\prime}_2}\left|\eta_3\right\rangle)\nonumber\\
+\frac{\alpha_3}{2}
(\left|10\right\rangle_{A_1A^{\prime}_1}\left|10\right\rangle_{A_2A^{\prime}_2}\left|\eta_0\right\rangle+
\left|10\right\rangle_{A_1A^{\prime}_1}\left|11\right\rangle_{A_2A^{\prime}_2}\left|\eta_1\right\rangle+\nonumber\\
\left|11\right\rangle_{A_1A^{\prime}_1}\left|10\right\rangle_{A_2A^{\prime}_2}\left|\eta_2\right\rangle+
\left|11\right\rangle_{A_1A^{\prime}_1}\left|11\right\rangle_{A_2A^{\prime}_2}\left|\eta_3\right\rangle),\nonumber\\
\end{eqnarray}
and the target state is
\begin{equation}
\left|\omega\right\rangle_{B_1B_2C_1C_2a_1a_2}\equiv\sum^3_{j=0}\alpha_j\left|\eta_j\right\rangle,
\end{equation}
which contains the optimal two clones of system $B_1B_2$ and
$C_1C_2$ respectively, as well as one ancilla of system $a_1a_2$
due to the universal Heisenberg QCM \cite{Cerf2}. Since either of
the senders $A_1$ and $A_2$ holds two particles being in the state
$\left|\Psi\right\rangle_{tot}$, they can individually perform a
joint measurement on its 2-qubit system in the Bell basis
\begin{eqnarray}
\left|\Phi^{\pm}\right\rangle=\frac{1}{\sqrt2}(\left|00\right\rangle\pm\left|11\right\rangle),\nonumber\\
\left|\Psi^{\pm}\right\rangle=\frac{1}{\sqrt2}(\left|01\right\rangle\pm\left|10\right\rangle).
\end{eqnarray}
Evidently, the resulting state is
$\langle\Phi^{\pm}|_{A_1A^{\prime}_1}\langle\Phi^{\pm}|_{A_2A^{\prime}_2}|\Psi\rangle_{tot}$,
etc, and there are in all 16 cases here. To simplify the
situation, we call the superscript ``$+$" or ``$-$" of the Bell
basis the parity of it. It is easy to show that any Bell
projection can be turned into one of the cases
$\langle\Phi^{\pm}|_{A_1A^{\prime}_1}\langle\Phi^{\pm}|_{A_2A^{\prime}_2}|\Psi\rangle_{tot}$
with the same parity as the former one, by using of the
state-transformations (9)-(12).

For example, if the measurement is taken in
$\{\left|\Phi^{-}\right\rangle_{A_1A^{\prime}_1},\left|\Psi^{-}\right\rangle_{A_2A^{\prime}_2}\}$,
the resulting state is
\begin{eqnarray}
\left|\Psi\right\rangle&=&\alpha_0\left|\eta_1\right\rangle-\alpha_1\left|\eta_0\right\rangle
-\alpha_2\left|\eta_3\right\rangle+\alpha_3\left|\eta_2\right\rangle.
\end{eqnarray}
By using of the state-transformation
$\left|\eta_0\right\rangle\leftrightarrow\left|\eta_1\right\rangle$
and
$\left|\eta_2\right\rangle\leftrightarrow\left|\eta_3\right\rangle$
(it requires the classical communication between the
participants), one can obtain
\begin{eqnarray}
\left|\Psi\right\rangle_{res}&=&\alpha_0\left|\eta_0\right\rangle-\alpha_1\left|\eta_1\right\rangle
-\alpha_2\left|\eta_2\right\rangle+\alpha_3\left|\eta_3\right\rangle,
\end{eqnarray}
which is the resulting state by measuring
$\left|\Psi\right\rangle_{tot}$ in
$\{\left|\Phi^{-}\right\rangle_{A_1A^{\prime}_1},\left|\Phi^{-}\right\rangle_{A_2A^{\prime}_2}\}$,
and its parity is unchanged. Similarly, one can check that the
resulting state derived from other Bell measurement can be turned
with the same identical parity by the state-transformation
operators. So it suffices to merely consider the cases of
measurements in
$\{\left|\Phi^{\pm}\right\rangle_{A_1A^{\prime}_1},\left|\Phi^{\pm}\right\rangle_{A_2A^{\prime}_2}\}$.
In particular, there are four subcases such that
$\{\left|\Phi^{+}\right\rangle_{A_1A^{\prime}_1},\left|\Phi^{+}\right\rangle_{A_2A^{\prime}_2}\},
\{\left|\Phi^{+}\right\rangle_{A_1A^{\prime}_1},\left|\Phi^{-}\right\rangle_{A_2A^{\prime}_2}\},\\
\{\left|\Phi^{-}\right\rangle_{A_1A^{\prime}_1},\left|\Phi^{+}\right\rangle_{A_2A^{\prime}_2}\}$
and $
\{\left|\Phi^{-}\right\rangle_{A_1A^{\prime}_1},\left|\Phi^{-}\right\rangle_{A_2A^{\prime}_2}\}$
here. Besides, the senders need broadcast the results of the
measurement to the receivers and ancillas so that they can perform
the unitary operations to modify the shared states locally. The
result by the measurement
$\{\left|\Phi^{+}\right\rangle_{A_1A^{\prime}_1},\left|\Phi^{+}\right\rangle_{A_2A^{\prime}_2}\}$
is precisely $\left|\omega\right\rangle_{B_1B_2C_1C_2a_1a_2}$. For
the second and third cases, by using of the parity-transformations
$\sigma_{zB_2}\otimes\sigma_{zC_2}\otimes\sigma_{za_2}$ and
$\sigma_{zB_1}\otimes\sigma_{zC_1}\otimes\sigma_{za_1}$
respectively, it is known from the relations (5)-(8) that the
receivers recover the correct state again. In case of the final
situation, it requires the collective rotations
$\sigma_{zB_1}\otimes\sigma_{zC_1}\otimes\sigma_{za_1}\otimes\sigma_{zB_2}\otimes\sigma_{zC_2}\otimes\sigma_{za_2}$
by all parties. Hence, one can always recover the target state and
thereby explicitly realize the optimal universal asymmetric
$1\rightarrow2$ telecloning of arbitrary two-qubit state by LOCC.

We investigate the scheme in terms of some figures of merit.
First, the required entanglement between senders and receivers is
$E(\left|\Omega\right\rangle_{A^{\prime}_1A^{\prime}_2B_1B_2C_1C_2a_1a_2})=2$
ebits. Besides, the classical cost informing the receivers and
ancillas is 4 cbits in all. Although the protocol in our paper is
sufficient to treat the optimal $1\rightarrow2$ asymmetric
telecloning of any 2-qubit input, the quantum cost here is not
always necessary for it has turned out that by using of only 1
ebit one can complete the optimal telecloning of a special family
of two-qubit states as described in the introduction \cite{Ghiu2}.
We readily prove that for the case of genuine 4-dimensional space,
namely $\alpha_0\alpha_1\alpha_2\alpha_3\neq0$, the cost of 2
ebits is also necessary for the telecloning scheme. Suppose that
the input state is maximally entangled with another qudit:
\begin{eqnarray}
\left|\psi^{\prime}\right\rangle_{A_1A_2A_3}=\frac12\big(\left|000\right\rangle+\left|011\right\rangle+
\left|102\right\rangle+\left|113\right\rangle\big).
\end{eqnarray}
Following the formal procedure described above, we can obtain the
resulting state
\begin{equation}
\left|\omega^{\prime}\right\rangle_{A_3B_1B_2C_1C_2a_1a_2}
=\sum^3_{j=0}\frac12\left|j\right\rangle\left|\eta\right\rangle_j.
\end{equation}
That is, the universal telecloning QCM of arbitrary 2-qubit state
can always create 2 ebits between the uncorrelated parties $A_3$
and the receivers. Since the entanglement cannot be increased on
average under LOCC \cite{Vedral}, we then assert that the cost of
2 ebits is always necessary and sufficient for this case. However,
for the case of $d=3$ namely there is a vanishing number among
$\alpha_0,\alpha_1,\alpha_2,\alpha_3$, it is difficult to show
that $\log_2 3$ ebits is the necessary amount of entanglement,
e.g., by a way similar to our scheme. A potentially feasible way
can be the $1\rightarrow M$ telecloning in \cite{Murao1}, but it
is necessary to find the decomposition of the unitary
transformations collectively performed on the system.

Second, the fidelity of our telecloning scheme is optimal. Due to
the optimal universal asymmetric Heisenberg QCM
\cite{Cerf2,Ghiu1}, for a $d$-level input state
$\left|\psi\right\rangle$ the clones have the form
\begin{equation}
\rho_B=[1+(d-1)(p^2+q^2)]^{-1}\big\{[1-q^2+(d-1)p^2]|\psi\rangle\langle\psi|
+q^2I\big\},
\end{equation}
and
\begin{equation}
\rho_C=[1+(d-1)(p^2+q^2)]^{-1}\big\{[1-p^2+(d-1)q^2]|\psi\rangle\langle\psi|
+p^2I\big\}.
\end{equation}
Then we can easily obtain the corresponding fidelities
\begin{eqnarray}
F_B(\rho_{\psi},\rho_B)=\frac{1+(d-1)p^2}{1+(d-1)(p^2+q^2)},\\
F_C(\rho_{\psi},\rho_C)=\frac{1+(d-1)q^2}{1+(d-1)(p^2+q^2)}.
\end{eqnarray}
Explicitly, they reach Werner's fidelity bound \cite{Werner} when
$p=q=1/2$. As our protocol derives from the case of $d=4$ of the
Heisenberg QCM, it is universal and not dependent on what the
input state is. Recently, N. Cerf \textit{et al.} \cite{Cerf3} has
proposed an optimal universal $1\rightarrow2$ QCM for maximally
entangled inputs. Their fidelity is a little higher than Werner's
bound, since the set of maximal entanglement is a small part of
the whole $d$-dimensional states. One can thus expect to get a
more efficient scheme of telecloning by following \cite{Cerf3}, as
well as other special QCMs such as the phase covariant cloning
\cite{Cinchetti} and real cloner \cite{Navez}, for both of them
contribute a higher fidelity than Werner's bound. However, as all
these potential schemes of telecloning are remarkably restricted
in the input states, our protocol gives a more universal plan. On
the other hand, it is difficult to create a better entanglement
QCM scheme for the unique characters it holds. As described by N.
Cerf \textit{et al.} \cite{Cerf3}, when the input state is
separable, the clones through the entanglement QCM should still be
separable. Moreover, such a protocol has to maximize the
entanglement of the clones, since it is regarded that some amounts
of entanglement of the initial state will lose during the cloning
process. Unfortunately, so far there is little progress for these
problems. The main difficulty originates in the mathematical
skills because of many variables in the deduction of optimal QCM,
and it is also hard to explore the asymmetric case \cite{Cerf3}.
Moreover, so far all QCMs of entanglement require the bipartite
inputs, which becomes fairly sophisticated if generalized to the
multipartite case. So it is difficult to create the telecloning
schemes by employing the universal QCMs of entangled states.
Comparatively speaking, we will show that our scheme can be
readily extended to the situation of multiqubit inputs and even
the mixed setting is also included.

Finally, we prove that our scheme does not create more
entanglement than that contained in the input state. The case of
maximally entangled input by the optimal QCM has been checked in
\cite{Cerf3}, i.e., when
$\mu\equiv|\alpha_0\alpha_3-\alpha_1\alpha_2|=1/2$. Here, we show
that this is a universal result for any $\mu$ of entangled input.
Due to the normalization condition of
$\left|\psi\right\rangle_{A_1A_2}$, we have $\mu\in[0,1/2]$. Let
\begin{eqnarray}
H(x)&\equiv&-(\frac12+\frac12\sqrt{1-x^2})\log_2(\frac12+\frac12\sqrt{1-x^2})\nonumber\\
&-&(\frac12-\frac12\sqrt{1-x^2})\log_2(\frac12-\frac12\sqrt{1-x^2}),
\end{eqnarray}
which is monotonically increasing with $x\in[0,1]$. One can simply
obtain the entanglement of the input state is
$E(\left|\psi\right\rangle_{A_1A_2})=H(2\mu)$. We employ the
entanglement of formation $E=H(C)$ \cite{Wootters}, where
$C=C_B(p)$ or $C_C(p)$ is the concurrence \cite{Coffman}, to
calculate the entanglement of the clones. Replace
$\left|\psi\right\rangle$ in $\rho_B$ with
$\alpha_0\left|00\right\rangle+\alpha_1\left|01\right\rangle+
\alpha_2\left|10\right\rangle+\alpha_3\left|11\right\rangle$, and
calculate the eigenvalues $\lambda_i$'s of
$\rho_B(\sigma_y\otimes\sigma_y)\rho^\ast_B(\sigma_y\otimes\sigma_y)$.
Notice
$F_B(\rho_{\psi},\rho_B)=F_B(p)=\frac{1+3p^2}{1+3(p^2+q^2)}$, some
simple algebra leads to
\begin{eqnarray}
C_B(p)&=&\mbox{max}\{0,\sqrt{\lambda_0}-\sqrt{\lambda_1}
-\sqrt{\lambda_2}-\sqrt{\lambda_3}\}\nonumber\\
&=&\mbox{max}\{0,(\frac83F_B-\frac23)\mu-\frac23(1-F_B)\},
\end{eqnarray}
where $\lambda_i$'s are decreasingly ordered. Similarly, let
$F_C(\rho_{\psi},\rho_C)=F_C(p)=\frac{1+3q^2}{1+3(p^2+q^2)}$ and
hence
\begin{eqnarray}
C_C(p)=\mbox{max}\{0,(\frac83F_C-\frac23)\mu-\frac23(1-F_C)\}.
\end{eqnarray}
Let $\Delta(\mu)=H(2\mu)-H(C_B(p))-H(C_C(p))$, then our assertion
is $\Delta(\mu)\geq0, \mu\in[0,1/2]$. We have analytically proven
it in appendix, so our scheme will never create more entanglement
than that contained in the original state. In addition, $F_B$ and
$F_C$ are not less than $1/4$ from their expressions. When one of
them reaches this lowest value, the other must be explicitly unit.
For the symmetric case namely $p=q=1/2$, we have $F_B=F_C=7/10$,
which reaches Werner's bound. Hence,
$C_B(1/2)=C_C(1/2)=\mbox{max}\{0,\frac65\mu-\frac15\}$ and the
maximal amount of entanglement created in either of the clones is
$H(C(\mu=1/2))=0.250225$ ebits. This is less than that in
\cite{Cerf3}, which is a special set of the two-qubit states.
Generally, the relation between entanglements created in the
clones constitute a teeterboard due to the monotonicity of
$H(C_B(p))$ and $H(C_C(p))$, i.e., if one of them decreases then
the other must increases, and vice versa.

\section{Optimal universal 1$\rightarrow$2 telecloning of n-qubit states
and 1$\rightarrow$4 telecloning of mixed states }

In this section we extend the 1$\rightarrow$2 telecloning to the
case of n-qubit pure states, and many properties of the above
scheme works here. Subsequently, we apply this universal scheme to
the $1\rightarrow4$ telecloning of arbitrary mixed state, which is
an interesting and difficult subject in QIT.

For convenience, we define the $n$-bit binary form of integer $N$.
Let $N=2^{n-1}\cdot c_{n-1}+\cdots+2^1\cdot c_1+2^0\cdot c_0$,
where $2^n>N$ and $c_i=0$ or 1, $\forall i.$ Then the unique
binary form is $\overline{N}=c_{n-1}\cdots c_1c_0$ ( we also write
$N=\overline{c_{n-1}\cdots c_1c_0}$ ). The situation here is that
$n$ separated senders $A_1,A_2,...,A_n$ share an arbitrary
multiqubit state
\begin{equation}
\left|\psi\right\rangle_{A_1,A_2,...,A_n}=\sum^{2^n-1}_{k=0}\alpha_k|\
\overline{k} \ \rangle_{A_1,A_2,...,A_n},
\end{equation}
where the coefficients $\alpha_i$'s satisfy
$\sum^{2^n-1}_{i=0}|\alpha_i|^2=1$, and the senders know nothing
about the state. They plan to optimally teleclone this state at
two remote locations, where two groups of uncorrelated receivers
$B_1,B_2,...,B_n$ and $C_1,C_2,...,C_n$ make up the system in the
clone respectively. Again, either of the participants in the whole
system can only operate locally and they can communicate with each
other. Consider the state $\left|\eta_j\right\rangle_{BC,anc}$ in
the last section. We write $j$ in its n-bit binary form and each
of the bit represents a party $B_i$ or $C_i$, namely
\begin{equation}
\left|\eta_j\right\rangle_{BC,anc}\sim\left|\eta_j\right\rangle_{B_1,B_2,...,B_n,C_1,C_2,...,C_n,a_1,a_2,...,a_n}.
\end{equation}
Having explained the form of $\left|\eta_j\right\rangle_{BC,anc}$,
we can propose the feasible quantum channel for the telecloning as
follows
\begin{equation}
\left|\Omega\right\rangle_{A^{\prime}BC,anc}=\frac{1}{2^n}\sum^{2^n-1}_{k=0}
\left|\overline{k}\right\rangle_{A^{\prime}_1,A^{\prime}_2,...,A^{\prime}_n}\left|\eta_k\right\rangle_{BC,anc}.
\end{equation}
Here, the particle $A^{\prime}_i$ belongs to the sender $A_i$. So
the total system is in the state
\begin{eqnarray}
\left|\Psi\right\rangle_{tot}&=&\left|\psi\right\rangle_{A_1,A_2,...,A_n}
\otimes\left|\Omega\right\rangle_{A^{\prime}BC,anc}\nonumber\\
&=&\sum^{2^n-1}_{k=0}\alpha_k|
\overline{k}\rangle_{A_1A_2,...A_n}\nonumber\\
&\otimes&\frac{1}{2^n}\sum^{2^n-1}_{k=0}
\left|\overline{k}\right\rangle_{A^{\prime}_1A^{\prime}_2...A^{\prime}_n}
\left|\eta_k\right\rangle_{BC,anc}\nonumber\\
&=&\frac{\alpha_0}{2^n}\Big( \sum^{2^n-1}_{k=0}
\left|\overline{0}\right\rangle_{A_1A_2...A_n}
\left|\overline{k}\right\rangle_{A^{\prime}_1A^{\prime}_2...A^{\prime}_n}
\left|\eta_k\right\rangle_{BC,anc}\Big)\nonumber\\
&+&\frac{\alpha_1}{2^n}\Big( \sum^{2^n-1}_{k=0}
\left|\overline{1}\right\rangle_{A_1A_2...A_n}
\left|\overline{k}\right\rangle_{A^{\prime}_1A^{\prime}_2...A^{\prime}_n}
\left|\eta_k\right\rangle_{BC,anc}\Big)\nonumber\\
\cdots\nonumber\\
&+&\frac{\alpha_{2^n-1}}{2^n}\Big( \sum^{2^n-1}_{k=0}
\left|\overline{2^n-1}\right\rangle_{A_1A_2...A_n}
\left|\overline{k}\right\rangle_{A^{\prime}_1A^{\prime}_2...A^{\prime}_n}\nonumber\\
&\otimes&\left|\eta_k\right\rangle_{BC,anc}\Big).
\end{eqnarray}

Next, the senders take measurement in the Bell basis on their
two-qubit systems respectively, and it is similar to that in the
telecloning of two-qubit states. The target state in the present
protocol is
\begin{equation}
\left|\Omega\right\rangle_{BC,anc}\equiv\sum^{2^n-1}_{j=0}\alpha_j\left|\eta_j\right\rangle,
\end{equation}
which contains two optimal asymmetric clones of system $B$ and $C$
\cite{Cerf2,Ghiu1}. However, we face a more complicated situation
here, and there are two main steps required for reaching the
target state, which also resembles two-qubit's case. First, we
prove that any Bell measurement can be turned into one of the
cases
$\langle\Phi^{\pm}|_{A_1A^{\prime}_1}\langle\Phi^{\pm}|_{A_2A^{\prime}_2}
...\langle\Phi^{\pm}|_{A_nA^{\prime}_n}|\Psi\rangle_{tot}$ with
the same parity as the former one by certain collective unitary
operations. That is, one can always obtain the state
\begin{equation}
\left|\Phi\right\rangle_{res}=\sum^{2^n-1}_{j=0}\alpha_j(-1)^{n_j}\left|\eta_j\right\rangle,
\end{equation}
where the sign $(-1)^{n_j}$ originates from the parity of Bell
projection. In order to get this result, we notice that every term
in $\left|\Psi\right\rangle_{tot}$ can be written as (the
coefficient is omitted)
\begin{equation}
T_{lk}=\left|a_1,a^{\prime}_1\right\rangle_{A_1,A^{\prime}_1}\left|a_2,a^{\prime}_2\right\rangle_{A_2,A^{\prime}_2}
\cdots\left|a_n,a^{\prime}_n\right\rangle_{A_n,A^{\prime}_n}\left|\eta_k\right\rangle,
\end{equation}
where $k=\overline{a^{\prime}_1a^{\prime}_2\cdots a^{\prime}_n}$,
and $l=\overline{a_1a_2\cdots a_n}$ denotes the sequence number of
$\alpha_l$ out of the bracket including this term. We call the
factor
$\left|a_m,a^{\prime}_m\right\rangle_{A_m,A^{\prime}_m},\forall m$
the secondary term of $T_{lk}$. Moreover, if $a_m=a^{\prime}_m$,
then this secondary term is $even$, otherwise it is $odd$.
Evidently, the even secondary term is the sum (or subtraction) of
the Bell basis $\left|\Phi^{\pm}\right\rangle$, while the odd one
is the sum (or subtraction) of the Bell basis
$\left|\Psi^{\pm}\right\rangle$. This implies that a single Bell
projection only operates on an even or odd secondary term.

Observe the terms in the $l$'th bracket in
$\left|\Psi\right\rangle_{tot},T_{l0},T_{l1},...$. One can find
that no two terms contain completely the same secondary terms with
respect to the position of every secondary term, since the
sequence number $l$ is unchanged. Hence, there must be a uniquely
residual term in every bracket after the Bell-measurement by the
senders. Denote $e_i$ the parity $``+"$ or $``-"$, and suppose the
measurement is taken in the sequence
$\{\left|\Psi^{e_{b_1}}\right\rangle_{A_{b_1}A^{\prime}_{b_1}},\cdots,
\left|\Psi^{e_{b_s}}\right\rangle_{A_{bs}A^{\prime}_{bs}}\}$, and
other two-qubit systems are projected onto the basis
$\{\left|\Phi^{e_m}\right\rangle_{A_mA^{\prime}_m}\}$. Such a
projection eliminates all but one term in every bracket, which has
$s$ odd secondary terms. Concretely for the $l$'th bracket, the
residual term is
\begin{eqnarray}
T_{lk}&=&\left|a_1,a_1\right\rangle_{A_1,A^{\prime}_1}\cdots
\left|a_{b_1},\widetilde{a_{b_1}}\right\rangle_{A_{b_1},A^{\prime}_{b_1}}\cdots
\left|a_{b_2},\widetilde{a_{b_2}}\right\rangle_{A_{b_2},A^{\prime}_{b_2}}\nonumber\\
&\cdots&\left|a_{b_s},\widetilde{a_{b_s}}\right\rangle_{A_{b_s},A^{\prime}_{b_s}}\cdots
\left|a_n,a_n\right\rangle_{A_n,A^{\prime}_n}\left|\eta_k\right\rangle,
\end{eqnarray}
where the tilde means the bit-shift,
$\widetilde{0}=1,\widetilde{1}=0.$ Thus
$k=\overline{a_1\cdots\widetilde{a_{b_1}}\cdots\widetilde{a_{b_2}}\cdots\widetilde{a_{b_s}}\cdots
a_n}$, and we must transform the term $\left|\eta_k\right\rangle$
into $\left|\eta_l\right\rangle$ for the $l$'th bracket
simultaneously. We can realize it by virtue of a collective
operation
$\prod^{s}_{i=1}\sigma_{xB_{bi}}\otimes\prod^{s}_{i=1}\sigma_{xC_{bi}}\otimes\prod^{s}_{i=1}\sigma_{xa_{bi}}$
acting on the state
$\left|\eta_j\right\rangle_{B_1,B_2,...,B_n,C_1,C_2,...,C_n,a_1,a_2,...,a_n}$.
This is similar to the state-transformation for two-qubit's case.
Therefore, the only term that operate in the $l$'th bracket is
$\left|\eta_l\right\rangle$, whose sign originates from
\begin{eqnarray}
&\prod^{n}_{i=1,\atop i\neq
b_1,b_2,...,b_s}&\left\langle\Phi^{e_i}|a_i,a_i\right\rangle_{A_i,A^{\prime}_i}\times
\prod^{s}_{i=1}\left\langle\Psi^{e_{b_i}}|a_{b_i},\widetilde{a_{b_i}}\right\rangle_{A_{b_i},A^{\prime}_{b_i}}\nonumber\\
&=&\prod^{n}_{i=1}\left\langle\Phi^{e_i}|a_i,a_i\right\rangle_{A_i,A^{\prime}_i},
\end{eqnarray}
where we have used the equation
$\left\langle\Psi^{e_{b_i}}|a_{b_i},\widetilde{a_{b_i}}\right\rangle_{A_{b_i},A^{\prime}_{b_i}}=
\left\langle\Phi^{e_{b_i}}|a_{b_i},a_{b_i}\right\rangle_{A_{b_i},A^{\prime}_{b_i}}$.
This means that an arbitrary Bell measurement on the state
$\left|\Psi\right\rangle_{tot}$ can be turned into the projection
$\langle\Phi^{e_1}|_{A_1A^{\prime}_1}\langle\Phi^{e_2}|_{A_2A^{\prime}_2}
...\langle\Phi^{e_n}|_{A_nA^{\prime}_n}|\Psi\rangle_{tot}$, while
the parity of each Bell projection is unchanged. In this process,
the senders need broadcast $2n$ classical bits to inform the
receivers of the results of measurement, so that the latter can
carry out the state-transformations.

Next, we focus on the sign of every term in
$\left|\Phi\right\rangle_{res}$, and this step is relatively
simpler. Suppose the sign solely originates in the projection of
some secondary term $\left|1,1\right\rangle_{A_k,A^{\prime}_k}$
(the term $\left|0,0\right\rangle_{A_k,A^{\prime}_k}$ never
contributes the sign), then it suffices to perform the operation
$\sigma_{zB_k}\otimes\sigma_{zC_k}\otimes\sigma_{za_k}$ on
$\left|\Phi\right\rangle_{res}$ to get the target state
$\left|\Omega\right\rangle_{BC,anc}$ (the classical communication
in the first step is available for this course). Generally, for
the sign produced by several projections by the senders
$A_{b_1},...,A_{b_s}$, one can recover the target state explicitly
by performing the unitary operation
$\prod^{s}_{i=1}\sigma_{zB_{bi}}\otimes\prod^{s}_{i=1}\sigma_{zC_{bi}}\otimes\prod^{s}_{i=1}\sigma_{za_{bi}}$.
The above operation hence can be regarded as a universal
parity-transformation.

In this scheme, the resource required is $n$ ebits and $2n$ cbits
in all between senders and receivers. Due to the property of
Heisenberg QCM, our scheme realizes the optimal unversal
asymmetric $1\rightarrow2$ telecloning. As any pure state can
always be composed of a certain number of qubits, we thus have
proposed a method to the telecloning of an arbitrary multipartite
state, while the expectant fidelity is also optimal due to
Werner's bound.

A useful application of this scheme is the telecloning of
multiqubit mixed states, which is an involved problem in the field
of quantum cloning \cite{Scarani}. An n-party mixed state can be
expressed as
\begin{equation}
\rho_{\psi}=\sum^{\sqrt
d-1}_{k=0}\alpha_k|\overline{k}\rangle_A\langle\overline{k}|,d=2^{2n}.
\end{equation}
Here, $\overline{k}$ also represents $n$ separated parties
$A_i,i=1,2,...,n$. On the other hand, we can regard $\rho_{\psi}$
as a reduced density operator of a pure multiqubit state
\begin{equation}
\left|\Psi\right\rangle=\sum^{\sqrt
d-1}_{k=0}\sqrt{\alpha_k}|\overline{k}\rangle_A|\overline{k}\rangle_{A^{\prime}}.
\end{equation}
Following the preceding scheme, we can obtain the asymmetric
telecloning of state $\left|\Psi\right\rangle$
\begin{equation}
\rho_{BB^{\prime}}=[1+(d-1)(p^2+q^2)]^{-1}\big\{[1-q^2+(d-1)p^2]|\Psi\rangle\langle\Psi|
+q^2I\big\},
\end{equation}
and
\begin{equation}
\rho_{CC^{\prime}}=[1+(d-1)(p^2+q^2)]^{-1}\big\{[1-p^2+(d-1)q^2]|\Psi\rangle\langle\Psi|
+p^2I\big\}.
\end{equation}
It should be pointed that although the expression of
$|\Psi\rangle$ in (38) and (39) is given by (37), the particles
held by the parties become $B,B^{\prime}$ and $C,C^{\prime}$
respectively. As the state $\left|\Psi\right\rangle$ is symmetric
with the exchange of system $B$ and $B^{\prime}$, so by tracing
out the freedom of $B^{\prime}$ one can get the clone of state
$\rho_{\psi}$
\begin{eqnarray}
\rho_B=[1+(d-1)(p^2+q^2)]^{-1}\nonumber\\
\big\{[1-q^2+(d-1)p^2]\sum^{\sqrt
d-1}_{k=0}\alpha_k|\overline{k}\rangle_B\langle\overline{k}|+\sqrt{d}q^2I\big\},
\end{eqnarray}
and $\rho_C$ is similarly recovered by $p\leftrightarrow q$.
Generally, this scheme indeed realizes the cloning of
$\rho_{\psi}$ to $\rho_B$, $\rho_C$, $\rho_{B^{\prime}}$ and
$\rho_{C^{\prime}}$. Thus it is a $1\rightarrow4$ asymmetric
telecloning of arbitrary mixed states, with the fidelities
$F(\rho_{\psi},\rho_{B})=F(\rho_{\psi},\rho_{B^{\prime}})$ and
$F(\rho_{\psi},\rho_{C})=F(\rho_{\psi},\rho_{C^{\prime}})$.
Moreover, both of them are higher than that of the pure
telecloning since the trace of a subsystem is a trace-preserving
quantum operation \cite{Chuang}, e.g.,
\begin{eqnarray}
F(\rho_{\psi},\rho_{B})&=&F(\mbox{Tr}[|\Psi\rangle\langle\Psi|],\mbox{Tr}[\rho_{BB^{\prime}}])\nonumber\\
&\geq&F(|\Psi\rangle\langle\Psi|,\rho_{BB^{\prime}}).
\end{eqnarray}
In the same way, we obtain $F(\rho_{\psi},\rho_{C})\geq
F(|\Psi\rangle\langle\Psi|,\rho_{CC^{\prime}})$. We explicitly
calculate this fidelity as a most figure of the scheme. The
fidelity of two mixed states is \cite{Jozsa}
\begin{equation}
F(\rho_1,\rho_2)=\Big[\mbox{Tr}\sqrt{\sqrt{\rho_1}\rho_2\sqrt{\rho_1}}\Big]^2.
\end{equation}
By some simple algebra, it follows that
\begin{eqnarray}
F(\rho_{\psi},\rho_{B})=[1+(d-1)(p^2+q^2)]^{-1}\nonumber\\
\Bigg(\sum^{\sqrt{d}-1}_{k=0}\sqrt{[1-q^2+(d-1)p^2]\alpha^2_k+\sqrt{d}q^2\alpha_k}\Bigg)^2.
\end{eqnarray}
Here, the unique restriction is
$\sum^{\sqrt{d}-1}_{k=0}\alpha_k=1,\forall{\alpha_k}\geq0$. By
employing the Lagrange multipliers, it is easy to prove that
$F(\rho_{\psi},\rho_{B})\in[\frac{1-q^2+(d-1)p^2+\sqrt d
q^2}{1+(d-1)(p^2+q^2)},1]$, and similarly
$F(\rho_{\psi},\rho_{C})\in[\frac{1-p^2+(d-1)q^2+\sqrt d
p^2}{1+(d-1)(p^2+q^2)},1]$. Comparing them with those for the pure
states, we find that the $1\rightarrow4$ asymmetric telecloning of
mixed states can be realized with a high fidelity by virtue of our
scheme.

\section{conclusions}

In this paper, we addressed the problem of asymmetric quantum
telecloning of arbitrary multipartite states in a universal case.
Our 1$\rightarrow$2 optimal scheme employed the Heisenberg QCM,
which explicitly reaches Werner's bound. We provided an important
application of this scheme on the 1$\rightarrow$4 universal
telecloning of mixed multiqubit states, with a fidelity higher
than that of the pure states. It is interesting that there may
exist some relations between the cloning of mixed states and
multipartite states. The present scheme cannot create more
entanglement than that of the original state. It is a problem to
extend our scheme to the case of 1$\rightarrow M$, so that the
entanglement can be remotely cloned more generally.

The work was partly supported by the NNSF of China Grant
No.90503009 and 973 Program Grant No.2005CB724508.

\begin{center}
{\bf APPENDIX}
\end{center}

Here we show that $\Delta(\mu)\geq0, \mu\in[0,1/2].$ For the case
of $\mu=1/2$,
$\Delta(1/2)=H(1)-H(\max\{0,2F_B-1\})-H(\max\{0,2F_C-1\})$. When
either of $2F_B-1$ and $2F_C-1$ is less than zero, the
monotonicity of $H(x)$ makes $\Delta(\mu)\geq0$. When both of them
are positive, it is easy to recover the assertion by plotting the
function $\Delta(1/2)$, whose independent variable is
$p\in[1/3,2/3]$. Moreover, when $\mu\in[0,1/6]$, there is at least
one zero in $C_B(p)$ and $C_C(p)$, and the monotonicity of $H(x)$
makes $\Delta(\mu)\geq0$ again. This implies that for the inputs
with $\mu\leq1/6$, it is impossible to create entanglement in both
of the clones simultaneously by our scheme. So it suffices to
investigate the case of $\mu\in(1/6,1/2)$, where both $C_B(p)$ and
$C_C(p)$ must be positive. Recall that $q=1-p$, we have
\begin{eqnarray*}
F_B(p)=\frac{1+3p^2}{4-6p+6p^2}>\frac{\mu+1}{4\mu+1},\\
F_C(p)=\frac{4-6p+3p^2}{4-6p+6p^2}>\frac{\mu+1}{4\mu+1},
\end{eqnarray*}
namely
\begin{equation*}
p\in\Big(\frac{1+\mu-\sqrt{4\mu+\mu^2}}{1-2\mu},\frac{-3\mu+\sqrt{4\mu+\mu^2}}{1-2\mu}\Big).
\end{equation*}
Mathematically, we only need calculate the derivative of
$\Delta(\mu)$ with respect to $p$, but it is difficult to do it in
this way because of the confused deduction. Notice that
$F_B(p)=F_C(1-p)$, so $H(C_B(p))=H(C_C(1-p))$, i.e., they are
symmetric and the symmetry axis is $p=1/2$. Thus we focus on the
property of $H(C_B(p))$. As $C_B(p)$ is monotonically increasing
with $p$, $H(C_B(p))$ is also monotonically increasing with $p$.
Calculate the second derivative of $H(C_B(p))$ with respect to
$p$,
\begin{eqnarray*}
\frac{\mathrm{d}^2}{\mathrm{d}p^2}H(C_B(p))&=&
\frac{\mathrm{d}}{\mathrm{d}p}\Big(\frac{\mathrm{d}H}{\mathrm{d}C}\frac{\mathrm{d}C}{\mathrm{d}p}\Big)\nonumber\\
&=&\frac{\mathrm{d}^2H}{\mathrm{d}C^2}\Big(\frac{\mathrm{d}C}{\mathrm{d}p}\Big)^2+
\frac{\mathrm{d}H}{\mathrm{d}C}\frac{\mathrm{d}^2C}{\mathrm{d}p^2}\nonumber\\
&=&\lambda\Big[\Big(\frac83\mu+\frac23\Big)\frac{2\sqrt{1-C^2}+\log_e\frac{1-\sqrt{1-C^2}}{1+\sqrt{1-C^2}}}
{C(1-C^2)\log_e\frac{1-\sqrt{1-C^2}}{1+\sqrt{1-C^2}}}\nonumber\\
&+&\frac{4(2-3p+3p^2)(5-9p-9p^2+9p^3)}{3(-1-2p+3p^2)^2}\Big],\nonumber\\
\end{eqnarray*}
where
$\lambda=\Big(\frac83\mu+\frac23\Big)\Big(\frac{\mathrm{d}F_B}{\mathrm{d}p}\Big)^2\frac{\mathrm{d}H}{\mathrm{d}C}$
is positive. As the first part and second part in the square
bracket are monotonically decreasing with $C=C_B(p)$ and $p$
respectively, $\frac{\mathrm{d}^2}{\mathrm{d}p^2}H(C_B(p))$ is
monotonically decreasing with $p$. By virtue of plotting it is
easy to show that
\begin{eqnarray*}
\frac{\mathrm{d}^2}{\mathrm{d}p^2}H(C_B(p))\bigg|_{p=0.56}>0,\nonumber\\
\frac{\mathrm{d}}{\mathrm{d}p}H(C_B(p))\bigg|_{p=2/3}>
\frac{\mathrm{d}}{\mathrm{d}p}H(C_B(p))\bigg|_{p=0.44}.
\end{eqnarray*}
Although the point $p=2/3$ is usually not in the physical region
$\Big(\frac{1+\mu-\sqrt{4\mu+\mu^2}}{1-2\mu},\frac{-3\mu+\sqrt{4\mu+\mu^2}}{1-2\mu}\Big)
$, the above argument mathematically applies to the region
$\Big(\frac{1+\mu-\sqrt{4\mu+\mu^2}}{1-2\mu},\frac23\Big]$. Thus
the inflection point of $H(C_B(p))$ is $p_{in}>0.56$. Consider the
sum of the $H(C_B(p))$ and $H(C_C(p))$, where
$H(C_B(p))=H(C_C(1-p))$. When
$p\in\Big[p_{in},\frac{-3\mu+\sqrt{4\mu+\mu^2}}{1-2\mu}\Big]$,
\begin{eqnarray*}
\frac{\mathrm{d}}{\mathrm{d}p}[H(C_B(p))+H(C_C(p))]>\nonumber\\
\frac{\mathrm{d}}{\mathrm{d}p}H(C_B(p))\bigg|_{p=2/3}-
\frac{\mathrm{d}}{\mathrm{d}p}H(C_B(p))\bigg|_{p=0.44}>0,
\end{eqnarray*}
and when $p\in\Big[1/2,p_{in}\Big]$, one readily obtains
$\frac{\mathrm{d}}{\mathrm{d}p}[H(C_B(p))+H(C_C(p))]>0$ as the
reflection point $p_{in}>0.56$. So $H(C_B(p))+H(C_C(p))$ is
monotonically increasing when
$p\in[1/2,\frac{-3\mu+\sqrt{4\mu+\mu^2}}{1-2\mu}]$. Due to the
symmetry of $H(C_B(p))$ and $H(C_C(p))$, the maximum of
$\Delta(\mu)$ is in the bound
$p=\frac{1+\mu-\sqrt{4\mu+\mu^2}}{1-2\mu}$ or
$\frac{-3\mu+\sqrt{4\mu+\mu^2}}{1-2\mu}$. This is just the case
where $C_B(p)$ or $C_C(p)$ vanishes, and thus $\Delta(\mu)\geq0$.
So we conclude that our scheme of 2-qubit telecloning will never
create more entanglement than that contained in the original
state.

\end{document}